\def\year{2021}\relax
\documentclass[letterpaper]{article} 
\usepackage{aaai21}  
\usepackage{times}  
\usepackage{helvet} 
\usepackage{courier}  
\usepackage[hyphens]{url}  
\usepackage{graphicx} 
\usepackage{natbib}  

\usepackage{caption} 
\usepackage[ruled,linesnumbered]{algorithm2e}
\SetKwComment{Comment}{$\triangleright$\ttfamily#1}{}
\usepackage{graphicx}
\usepackage{subcaption}
\usepackage{algorithmic}
\algsetup{linenosize=\small}
\usepackage{todonotes}
\usepackage{mathtools}
\SetCommentSty{mycommfont}
\let\oldnl\nl
\newcommand{\nonl}{\renewcommand{\nl}{\let\nl\oldnl}}
\RequirePackage[rightcaption]{sidecap}
\RequirePackage{enumitem} \setlist{nosep}
\RequirePackage{soul}
\RequirePackage{bm}
\usepackage{amssymb}
\usepackage{pdfrender}
\newcommand*{\boldcheckmark}{%
  \textpdfrender{
    TextRenderingMode=FillStroke,
    LineWidth=.5pt, 
  }{\checkmark}%
}
\SetKwComment{Comment}{$\triangleright$\ }{}
\newcommand{\graphname}{\texttt{CCN}}
\newcommand{\wicci}{\texttt{WICCI}}
\newcommand{\algonamegraph}{\texttt{KORSE}}
\newcommand{\algonameml}{\texttt{NURSE}}
\title{Weakening the Inner Strength: Spotting  Core Collusive Users in YouTube Blackmarket Network}

\author {
    Hridoy Sankar Dutta$^*$, Nirav Diwan\thanks{Both authors contributed equally to the paper.} and
    Tanmoy Chakraborty \\
}

\affiliations {
    Dept of Computer Science, IIIT Delhi, India \\
    hridoyd@iiitd.ac.in, nirav17072@iiitd.ac.in, tanmoy@iiitd.ac.in
}
\begin{document}

\maketitle
\begin{abstract}
Social reputation  (e.g., likes, comments, shares, etc.) on YouTube is the primary tenet to popularize channels/videos. However, the organic way to improve social reputation is tedious, which often provokes content creators to seek services of online blackmarkets for rapidly inflating content reputation. Such blackmarkets act underneath a thriving collusive ecosystem comprising {\em core users} and {\em compromised accounts} (together known as {\em collusive users}). Core users form the backbone of blackmarkets; thus, spotting and suspending them may help in destabilizing the entire collusive network.   Although a few studies focused on collusive user detection on Twitter, Facebook, and YouTube, none of them differentiate between core users and compromised accounts. 
  
   We are the first to present a rigorous analysis of core users in YouTube blackmarkets. 
   To this end, we collect a new dataset of collusive YouTube users. We study the {\em core-periphery structure} of the underlying {\em collusive commenting network} (\graphname). We examine the topology of \graphname\ to explore the behavioral dynamics of core and compromised users. We then introduce \algonamegraph, a novel graph-based method to automatically detect core users based {\em only} on the topological structure of \graphname. \algonamegraph\ performs a weighted $k$-core decomposition using our proposed metric, called {\em Weighted Internal Core Collusive Index} (\wicci). However, \algonamegraph\ is infeasible to adopt in practice as it requires complete interactions among collusive users to construct \graphname.
   We, therefore, propose \algonameml, a deep fusion framework that {\em only leverages user timelines} (without considering the underlying \graphname) to detect  core blackmarket users. Experimental results  show that \algonameml~is quite close to \algonamegraph\ in detecting core  users and outperforms nine baselines.
\end{abstract}

\section{Introduction}
In recent years, YouTube has grown as a primary video-sharing platform, where content creators create channels and upload videos. The videos are then recommended to the content consumers based on several factors, one of which is the online {\em social reputation} of the creators and their content. Social reputation is usually quantified by the endorsement of the viewers in terms of likes, (positive) comments, shares, etc. However, an organic way of gaining reputation is a time consuming process, and often depends on several other factors such as the quality and relevance of the video, initial viewers and their underlying connections. Unfortunately, there exist a handful of online reputation manipulation services ({\em aka} blackmarkets) which help content creators rapidly inflate their reputations in an artificial way \cite{shah2017many}. Such services are built on a large thriving ecosystem of collusive network. The underlying network comprising {\em core users} -- fake accounts or sockpuppets \cite{bu2013sock},  which are fully controlled by the blackmarkets (puppet masters), and {\em compromised accounts} which are temporarily hired to support the core users --  these two types of users are together called as {\em collusive users}. Core users are the spine of any collusive blackmarket; they  monitor and intelligently control the entire fraudulent activities in such a way that none of their hired compromised accounts are suspended. Therefore, detecting and removing core blackmarket users from YouTube is of utmost importance to decentralize the collusive network and keep the YouTube ecosystem healthy and trustworthy. In this study, we deal with {\em freemium blackmarkets} \cite{shah2017many} which invite customers to opt for the service for free, in lieu of surrendering their accounts temporarily for blackmarket activities. In doing so, customers gain virtual credit and use it to grow their content's reputation.


\begin{figure}[!t]
    \centerline{\includegraphics[width=0.99\columnwidth]{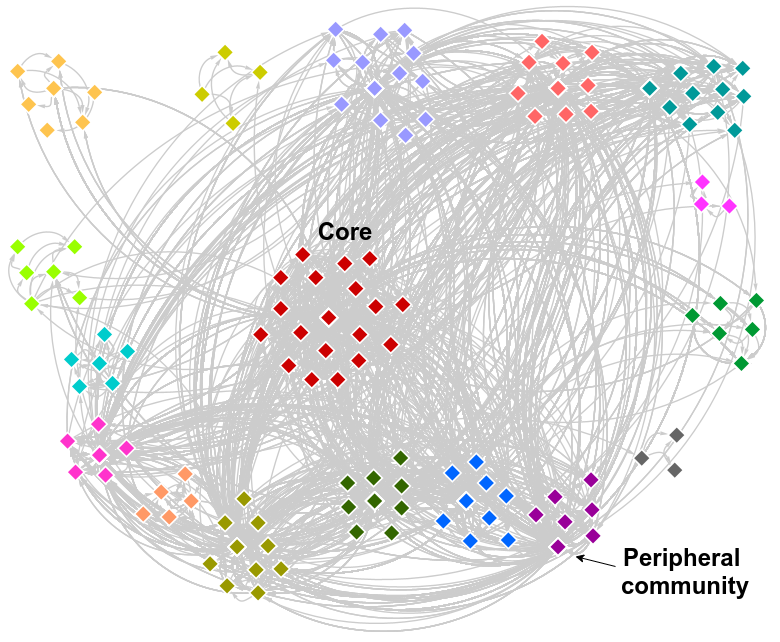}}
    \caption{Visualization of the collusive commenting network (\graphname). Unlike conventional core-periphery structure where peripheral nodes are sparsely connected internally, \graphname\ constitutes dense peripheral communities sparsely connected with the core, indicating the growth of the network up to a certain point where it may not require  core users to support compromised users for self-sustainability.}
  \label{fig:demo}
\end{figure}

{\bf State-of-the-art and Motivation.}
Several efforts have been made to detect {\em fake activities} in different online social networks \cite{cresci2015fame,castellini2017fake}. However, as suggested by \citet{dutta2020detecting}, {\em collusive activities} are very different from usual fake activities. A few  studies attempted to explore the dynamics of blackmarkets, mostly for Twitter \cite{castellini2017fake,dutta2019blackmarket} and Facebook \cite{farooqi2017measuring}.
On YouTube, there exists only one method, named {\tt CollATe} to detect collusive users  \cite{dutta2020detecting}. However, to our knowledge, none of these methods attempted to further divide collusive users into  core  and compromised accounts. 

\begin{table}[!t]
    \centering
    \caption{Important notations and denotations.}
    \label{tab:notation}
    \scalebox{0.9}{
    \begin{tabular}{p{0.28\columnwidth}|p{0.7\columnwidth}}
    \hline
    
         {\bf Notation} & {\bf Denotation}  \\\hline 
        $G(N,E)$ & Collusive commenting network\\
        $V$ & Set of sets where $\{v_i\}\in \bm{V}$ indicates the set of videos created and posted by user $n_i$\\
        $v_{i,j}$ & $j^{th}$ video in the video set $\bm{v_{i}}$\\
        $comments(n,c)$ & No. of comments posted by user $n$ on video $c$\\
        $w_{ij}$ & Weight of the edge connecting nodes $n_i$ and $n_j$\\
        $wc$ & weighted coreness score\\
        $core_{th}$ & Coreness threshold\\
        $G_C$ & Core subgraph\\
        $G_{P}$ & Induced subgraph of the peripheral nodes\\
        $G_{P}^{L}$ & Largest connected component in $G_P$\\
        $WCS_{Core,C}$ & Weighted cut set between the core and a peripheral community $C$\\\hline
    \end{tabular}}
\end{table}
One may argue that once a collusive account (be it core or compromised) is detected, it should be banned. Then why do we need to explicitly identify core and compromised accounts, while both of them deserve punishment? We argue that the role of a core user is different from a compromised account in the collusive ecosystem; therefore, the extent of punishment may differ. Compromised users are more interested in self-promotion; they join blackmarkets temporarily; they gain appraisals for their online content both organically (genuine interest by other users) and inorganically (through blackmarket services). However, core users, being the backbone of the blackmarkets, always intend to grow and popularize their business. They are permanent members of the blackmarkets; they provoke other users to join the services; and they generally initiate the artificial inflation of the reputation of online content. Therefore, they are more harmful to pollute the online ecosystem. Due to such contrasting behavior of core and compromised users, one may consider that core users should be punished differently than compromised users. For instance, a complete ban of core users would limit the growth of the collusive blackmarkets. However, for compromised users, it may be wise to just warn them and restrict their social network activities for a limited time, instead of a complete ban. The authorities of a social media platform may design suitable policies to handle these two cases.

To our knowledge, \textbf{ours is the first attempt to identify and explore the dynamics of \textit{core blackmarket users}. It is also the second attempt after  {\tt CollATe} \cite{dutta2020detecting} to explore \textit{YouTube blackmarkets}.}

{\bf Present Work: \algonamegraph.} In this paper, we investigate the dynamics of core users in YouLikeHits, one of the popular YouTube blackmarket services. We start by collecting a novel dataset from YouLikeHits and YouTube, consisting of collusive users, the videos they promote through blackmarkets, and their comments on YouTube videos. In this study, we deal with only one type of appraisals i.e., {\em collusive comment} on YouTube videos. We then construct a collusive commenting network (\graphname) based on the co-commenting activities among collusive users. We leverage the topological structure of \graphname\ to detect core users using our proposed method, \algonamegraph\ which utilizes $k$-core decomposition particularly designed based on our proposed metric, {\em Weighted Internal Core Collusive Index} (\wicci).

{\bf Present Work: Core-periphery Structure.} An exhaustive analysis on the interactions of core and peripheral nodes reveals a counter-intuitive core-periphery structure of \graphname\ -- unlike a conventional network where peripheral nodes are sparsely connected, and get disconnected upon removal of the core,  \graphname\ constitutes peripheral nodes which form several small and dense communities around the core (c.f. Fig. \ref{fig:demo}). We further observe that there exists a strong positive correlation between the internal interactions within peripheral communities and the interactions between the core and the peripheral communities. This gives us the evidence that in peripheral communities, compromised users  who comment heavily on videos that are co-commented by core users,  tend to contribute more to the collusive market.
We also present a case study  to highlight the major differences between core and compromised users  based on their user timelines: (i) Core users, although act as heavy contributors of the blackmarket services, are not the top beneficiaries of the collusive market. (ii) Core users indulge in less self-promotion of videos. (iii) Core users are less active participants of the collusive market than compromised users; they initiate the fraudulent activities and let the compromised users finish the remaining job.

{\bf Present Work: \algonameml.} Although \algonamegraph\ is highly accurate in detecting core users, it is practically infeasible to deploy as it requires the complete snaphot of the collusive market on a streaming basis and is also required to be re-run on the introduction of each new user. Therefore, we consider core users detected by \algonamegraph\ as the ground-truth\footnote{Collecting the ground-truth for fake/genuine entity detection is challenging, which  usually requires annotations from annotators with domain expertise \cite{shu2018understanding}. However, obtaining the ground-truth data of core blackmarket users is almost impossible. We do not know any legal way to find ``core'' blackmarket users. Therefore, we consider \algonamegraph\ as an oracle, which cannot be used in practice but can be used to create the ground-truth.  One can argue that the current way of creating the ground-truth may be unconvincing.  However, we perform several case studies to provide strong empirical evidence which may validate our strategy of collecting the ground-truth.
We do not know any other way of ground-truth creation for this problem unless blackmarkets themselves provide the same!} and develop \algonameml, a deep fusion framework that {\em only} considers user timeline (without the underlying \graphname) and video submission information to detect core blackmarket users. Experiments on our curated dataset show that \algonameml\ is quite close to \algonamegraph\ with $0.879$ F1-Score and $0.928$ AUC, outperforming nine baselines. 

{\bf Contributions:} In short, our contributions are four-fold:
\begin{itemize}
    \item {\bf Novel problem:} We are the first to address the problem of {\em core blackmarket user} detection.
    \item {\bf Unique dataset:} Our curated dataset is the first dataset, comprising {\em core and compromised collusive YouTube users}.
    \item {\bf Novel methods:} Our proposed methods, \algonamegraph\ and \algonameml, are the first in detecting core blackmarket users.
    \item {\bf Non-intuitive findings:} Empirical analysis of the dynamics of core and compromised users reveals several non-trivial characteristics of blackmarket services.
\end{itemize}
{\bf Reproducibility.} Our full code  and dataset are available here -
https://github.com/LCS2-IIITD/ICWSM-2022-Core-Collusive-Youtube-BlackMarket


\section{Related Work}
We summarize related studies by dividing them in two subsections: (i) blackmarkets and collusion, and (i) network core detection.

\textbf{Blackmarkets and Collusion:} 
Recently, the activities of blackmarket services have garnered significant attention among the researchers due to the way they provide artificial appraisals to online media content. \citet{shah2017many} provided a broad overview of the working of blackmarkets. 
\citet{dutta2019blackmarket} attempted to detect collusive retweeters on Twitter. The authors also mentioned how collusive users are asynchronous in nature as compared to normal retweet fraudsters.
\citet{dutta2019blackmarket} further studied the working of premium and freemium blackmarket services in  providing collusive appraisals on Twitter. 
\citet{arora2019multitask} proposed a multitask learning framework to detect tweets submitted to blackmarkets for collusive retweet appraisals. \citet{arora2020analyzing} further investigated the blackmarket customers engaged in collusive retweeting activities using a multiview learning based approach. 
\citet{chetan2019corerank} proposed \texttt{CoReRank}, an unsupervised method to detect collusive retweeters and suspicious tweets on Twitter. 
\citet{farooqi2019measurement} proposed the measurement and early detection of third-party application abuse on Twitter.
\citet{farooqi2017measuring} showed how collusion networks collect OAuth access tokens from colluding members and abuse them to provide fake likes or comments to their members.
\citet{zhu2016new} proposed an automated approach to detect collusive behavior in question-answering systems.
\citet{dhawan2019spotting} proposed \texttt{DeFrauder}, an unsupervised framework to detect collusive behavior of online fraud groups in customer reviews. 
Several other studies focused on detecting fake followers on Twitter \cite{cresci2015fame,castellini2017fake}, fake likes on Instagram \cite{sen2018worth} and fake  views in video-sharing platforms \cite{shah2017flock}. 
\citet{dutta2020detecting} is the closest to the current research, which detects collusive blackmarket users on YouTube. However, it does not focus on detecting {\em core blackmarket users}. 
\begin{table}[!t]
 \caption{Qualitative comparison of \algonamegraph\ and \algonameml~with similar approaches.}
 \label{table:comparison_related_work}
 \centering
 \scalebox{0.8}{
 \begin{tabular}{l|l l l l l l| l l}
    & \rotatebox{90}{\citet{batagelj2003m}} & \rotatebox{90}{\citet{shin2016corescope}} & \rotatebox{90}{\citet{cheng2011efficient}}  & \rotatebox{90}{\citet{rombach2014core}}  & \rotatebox{90}{\citet{zhang2017finding}} & \rotatebox{90}{\citet{dutta2020detecting}}  
     & \rotatebox{90}{\algonamegraph} & \rotatebox{90}{\algonameml}  \\ \hline
Detect collusive users &   &   &   & &    & \checkmark &$\boldcheckmark$ & $\boldcheckmark$\\ 
Detect core blackmarket users &  &  & &  &  &   &$\boldcheckmark$&$\boldcheckmark$ \\
Graph-based approach & \checkmark & \checkmark   & \checkmark &  \checkmark &  \checkmark &   &$\boldcheckmark$& \\ 
Deal with weighted graph & \checkmark &  &    &  \checkmark &    & &$\boldcheckmark$ & \\
Consider profile information &  &  &   &    &    &   \checkmark &  &$\boldcheckmark$\\ 
Consider content information &  &  &   &   &    & \checkmark&    &$\boldcheckmark$ \\ \hline
 \end{tabular}}
\end{table}

\textbf{Network Core Detection:}  
 Due to the abundance of literature on network core detection, we restrict our discussion to some selected works that we deem as pertinent to our study. $k$-core decomposition \cite{batagelj2003m} is considered to be the {\em de facto} to detect core nodes. It is based on the recursive removal of vertices that have degree less than $k$ in the input network.  \citet{rombach2014core} proposed an algorithm to detect core-periphery structure in networks. The goal of this algorithm is to  identify densely connected core nodes and sparsely connected peripheral nodes. 
 \citet{cucuringu2016detection} detected core and periphery using spectral methods and geodesic paths. 
 \citet{kojaku2017finding} discovered multiple non-overlapping groups of core-periphery structure by maximizing a novel quality function which compares the number of edges of different types in a network. \cite{xiang2018unified} detected multiple core-periphery structures and communities based on network density. The authors also proposed an improved version of their model to detect active and overlapping nodes. 
 \citet{zhang2017finding} studied the problem of collapsed $k$-core to identify a set of vertices whose removal can lead to the smallest $k$-core in the network. \citet{shin2016corescope} showed empirical patterns in real-world graphs  related to $k$-cores. \citet{laine2011user} explored the dynamics of social activities and communities in the context of grouping behaviors on YouTube.
Recently, it has been observed that the subgraphs of the detected core users are used for several graph-related tasks, such as community detection \cite{peng2014accelerating,xiang2018unified}, dense-subgraph detection \cite{andersen2009finding,hooi2020telltail}, and graph visualization \cite{alvarez2005k}. 
We encourage the readers to go through \citet{malliaros2016core} for a comprehensive survey on network core detection.


 \textbf{Differences with Existing Studies:} Table \ref{table:comparison_related_work} compares our methods (\algonamegraph~and~\algonameml) with a few relevant studies. In short, our methods are different from others in five aspects -- (i) we are the first to address {\bf core blackmarket user detection} problem; (ii) we are the second after \cite{dutta2020detecting} to deal with {\bf YouTube} collusive blackmarkets; (iii) We propose {\bf both unsupervised} (\algonamegraph) and {\bf supervised} (\algonameml) methods for core detection; (iv) our {\bf dataset comprising  core blackmarket users} is unique; and (v) we provide a {\bf rigorous analysis} to explore the dynamic of core and compromised users.

\section{Methodology}
\subsection{Dataset Description} \label{sec:dataset}

In this work, we consider YouLikeHits\footnote{\url{https://www.youlikehits.com/}}, a freemium blackmarket service\footnote{Freemium blackmarkets offer customers to enjoy their services for free with the condition that the customers will temporarily act on behalf of the blackmarkets. Upon signing up, the social media accounts of customers are compromised for a limited time for blackmarket activities, which in turn help them gain virtual credits}. 
We designed web scrapers to extract the ids of YouTube videos submitted to blackmarket services for collusive comments. We used YouTube API\footnote{\url{https://developers.google.com/youtube/v3}} to extract the metadata details and comment history of these videos. We extracted $26,166$  YouTube videos which were submitted to YouLikeHits for collusive comments. These videos were uploaded to $11,000$ unique YouTube channels. To our knowledge, this is the first dataset of its kind. Note that the entire data collection process was performed after taking proper Institutional Review Board (IRB) approval.

\begin{figure}[!h]
    \centering
    \captionsetup[subfigure]{labelformat=empty}
    \subfloat{{\includegraphics[width=0.49\columnwidth]{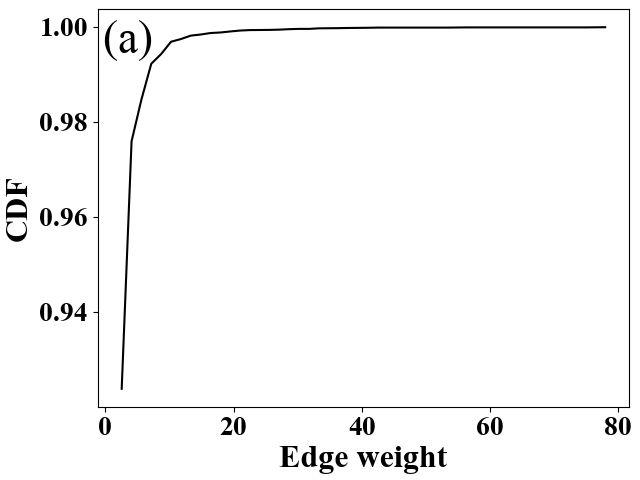}}}
    \subfloat{{\includegraphics[width=0.49\columnwidth]{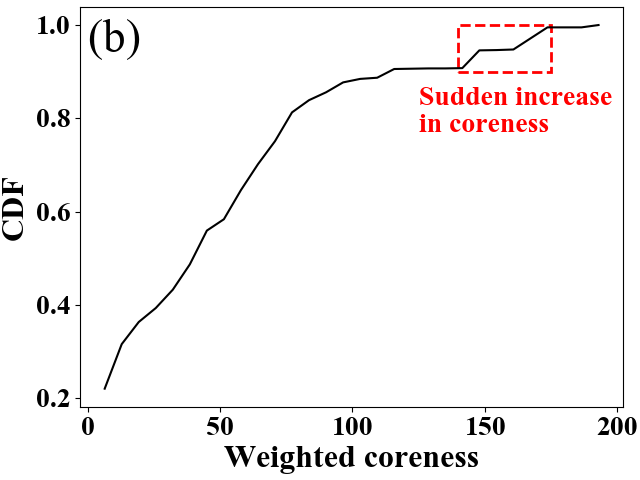}}}
    \caption{Cumulative distribution of (a) edge weights, and (b) weighted coreness scores of nodes in  \graphname. Contrary to the general observation that coreness score follows power law, we observe that there are relatively large number of nodes having  high weighted coreness.}
    \label{fig:coreness}
\end{figure}

\subsection{Preliminaries and Graph Construction}
Here we present some important concepts used throughout the paper. 
Table \ref{tab:notation} summarises important notations.


[\textbf{Collusive Users and Videos}]
We define \textit{collusive users} as those who are involved in the blackmarket activities. There are two types of collusive users -- core users and compromised users.  
We call the videos submitted to freemium blackmarkets as collusive videos.

[\textbf{Core Users}]
A limited set of online accounts are fully controlled by the blackmarket authorities. These accounts can be bots (fully automated), sockpuppets (controlled by puppet masters) \cite{bu2013sock} or fake accounts. However, they are used only to benefit blackmarkets. We call these users core blackmarket users. 

[\textbf{Compromised Users}]
These are YouTube content creators who submit their content to the freemium blackmarkets in order to receive artificial comments within a short duration. Being freemium customers, their accounts are compromised for a limited time to  perform illegal activities by commenting on videos of other blackmarket customers.

[\textbf{Collusive Commenting Network (\graphname)}]
A \graphname~is an undirected and weighted network $G(N,E)$, where each node $n\in N$ represents a collusive user, and two nodes $n_i$ and $n_j$ are connected by an edge $e_{ij}=\langle n_i, n_j\rangle$ if the corresponding users co-commented on the same videos. The weight $w_{ij}$ of the edge $e_{ij}$ is calculated as per Eq. \ref{eq:weight}.

Let us denote a set of sets, $V = \{\{\bm{v_1}\},\{\bm{v_2}\},\{\bm{v_3}\},\dots\}$, where $\{\bm{v_i}\}$ indicates the set of videos posted by collusive user $n_i$. $\{\bm{v_{i,j}}\}$ indicates the $j^{th}$ video in the set $\bm{v_i}$.



[\textbf{Inter-user Comment Count}]
The number of comments posted by the collusive user $n$ on video $c$ is denoted by $comments(n,c)$. We define \textit{Inter-user comment count} (IUCC) for a video $c$ and a pair of users $n_i$ and $n_j$ as the minimum of the number of comments by $n_i$ and $n_j$ on $c$.
\begin{equation}\small\label{eq:weight}
IUCC(n_1,n_2,c) = min\big(comments(n_1,c),comments(n_2,c)\big)
\end{equation}

[\textbf{Edge weight}]\label{def;weight}
We measure the \textit{edge weight} between two nodes (collusive users) $n_i$ and $n_j$ as follows:

\begin{equation}\label{eq:IUCC}
w_{ij} = \sum_{\substack{p = 1\\p \neq i,j}}^{|V|}\hspace{1mm} \sum_{q = 1}^{|v_p|} IUCC(n_{i},n_{j},v_{p,q})
\end{equation}

The edge weight $w_{ij}$ indicates the aggregated IUCC across all the videos co-commented by $n_i$ and $n_j$, excluding their own videos. We exclude the videos created by $n_i$ and $n_j$ since the comments on these videos can be easily manipulated (added or deleted) by the owners themselves. Table \ref{tab:summaryCCN} summarises the properties of \graphname. Fig. \ref{fig:coreness}(a) shows the cumulative distribution of $w_{ij}$.



\begin{table}[!t]
    \centering
    \caption{Topological properties of \graphname.}
    \label{tab:summaryCCN}
    \begin{tabular}{l|c}\hline
         \multicolumn{1}{c|}{\bf Property} & {\bf Value}  \\\hline
         \# nodes  & $1,603$ \\
         \# edges  & $51,424$\\
         Avg./max/min edge weight & $1.392$ / $78$ / $1$\\
         Avg./max/min weighted degree of nodes & $89.367$ / $1638$ / $1$\\
         Unweighted edge density & $0.040$\\
         Unweighted clustering coefficient &$0.737$ \\
         Network diameter & 8\\\hline
    \end{tabular}
\end{table}

\subsection{Weighted $k$-core Decomposition}
Given a graph $G(N,E)$, the weighted $k$-core detection problem aims to find \textit{k-core} (or core of order $k$), the maximal induced subgraph  denoted by $G_k(N_k,E_k)$ such that $G_k \subseteq G$ and $\forall n \in N_k: deg(n) \ge k$.
The following two methods are often used to solve this problem: \textit{k-core decomposition} \cite{rombach2014core} and \textit{core-periphery algorithm} \cite{della2013profiling}. In our case, we choose \textit{k-core decomposition}\footnote{We use the weighted version of \textit{k-core decomposition} to incorporate the edge weights (see Eq. \ref{eq:weight} for more details).}. In (weighted) k-core decomposition, to detect core users, we repeatedly delete nodes with (weighted) degree\footnote{The weighted degree of a node is the sum over the edge weights of the connected edges.} less than $k$ until no such node is left (this is also known as ``shaving'' method \cite{shin2016corescope}). The reasons behind choosing $k$-core decomposition are as follows: (i) It has been empirically shown to be successful in modeling user engagement \cite{zhang2016engagement, zhang2017finding}; (ii) Unlike $k$-core, core-periphery algorithm fits more closely with networks where the nodes are not closely connected to each other \cite{borgatti2000models}. However, in blackmarket services, the sole purpose of collusive users to join the services is to gain credits (by providing collusive appraisals to the content of other users) which can be used by them to artificially inflate their social growth. This strengthens the connectivity among the collusive users.
The reason behind expecting high interactions among users stems from the fact highlighted in \cite{dutta2020detecting} that different collusive users retweet the same tweets on the collusive market regardless of the topic of the tweets. We expect a similar behavior in case of YouTube comments, i.e., different collusive users tend to comment on the same videos in order to earn credits. In our dataset, a collusive video has an average of  $3$ comments by collusive users. This would create more relations (edges) between nodes in \graphname. 

\subsection{\wicci: Expected Behavior of Core Users}

We frame the core detection problem in \graphname~as the weighted $k$-core decomposition problem in \graphname. 
$k$-core decomposition assigns a \textit{coreness value} to each vertex. In our case, the coreness value ranges from $1$ to $193$, with an average value of $48.7$. We obtain an ordered list of vertices sorted in decreasing order of the coreness value. Typically, the node assigned with the highest coreness value is said to be the ``most influential node'' in the graph. The subgraph formed with such highly influential vertices is known as \textit{degeneracy-core} or $k_{max}$-core. On running the weighted $k$-core decomposition on  \graphname, we obtain a \textit{degeneracy-core} consisting of $8$ users. We expect the distribution of nodes to continually decrease with increasing coreness, as observed in typical core-periphery structures. However, we observe that the fraction of nodes with a high weighted coreness is unusually high ($~$ 12.1\% users with $\geq 100$ coreness score as shown in Fig. \ref{fig:coreness}(b)). This indicates the presence of {\em a larger set of core users}.

Therefore, in \graphname, to define the partition of core and compromised users, we propose a metric, called \textit{Weighted Internal Core Collusive Index} (\wicci) which is motivated by 
\citet{rombach2014core}. \wicci\ is used to partition the  list of decreasing weighted coreness values by a ``coreness threshold''. The nodes whose coreness is above the threshold are eligible to be the core nodes, while the remaining nodes are considered as compromised users. 
To define \wicci, we consider two important properties of core users as follows:

\begin{enumerate}
    \item \textbf{Density:} 
    A core component of a network should be densely connected \cite{rombach2014core,borgatti2000models}. 
    We attempt to understand the implications of a dense core in  \graphname, by considering the flip-side first -- {\em a sparse core}. A sparse core in \graphname~would have less number of edges connecting vertices internally. In the current scenario, it implies that different users have commented upon different sets of videos. However, the existence of such an entity would mean that there is no cohesion or strategy  in the way core users operate. They may be commenting randomly on different videos. The existence of a dense core, however, would imply that different users are commenting on a same set of (collusive) videos, indicating some cohesion or strategy.
    Note that when we increase the coreness threshold, the subgraph of the core formed has an increasing density (and a decreasing size).
    \begin{equation}
    \wicci \propto density^{\beta} 
    \end{equation}
    where $\beta$ is the density coefficient. We utilize $\beta$ to vary the proportionality of \wicci~with density.
    
    \item \textbf{Fraction of weighted size of core:} There is a major flaw in considering only density to define a core. Density does not take into account the edge weight i.e., the volume with which the two users have commented together on same videos. We intuitively expect that inside a core, a high fraction of the commenting activities take place.  We define $W_G$ as the weighted size (sum of the weights of edges) of  \graphname~and $W_C$ as the weighted size of the core subgraph $G_C$. Correspondingly,
    
    \begin{equation}\label{eq:weighted1}
        \wicci \propto \frac{W_C}{W_G}
    \end{equation}
    
    Combining (3) and (4), we get
    \begin{equation}\label{eq:weighted2}
        \wicci = k \times  \frac{W_C}{W_G} \times density^{\beta}
    \end{equation}
    where $k$ is the constant of proportionality. We assume it to be $1$.


    
\end{enumerate}

\begin{figure*}[!t]
    \centering
    \captionsetup[subfigure]{labelformat=empty}
    \subfloat[]{{\includegraphics[width=5.6cm]{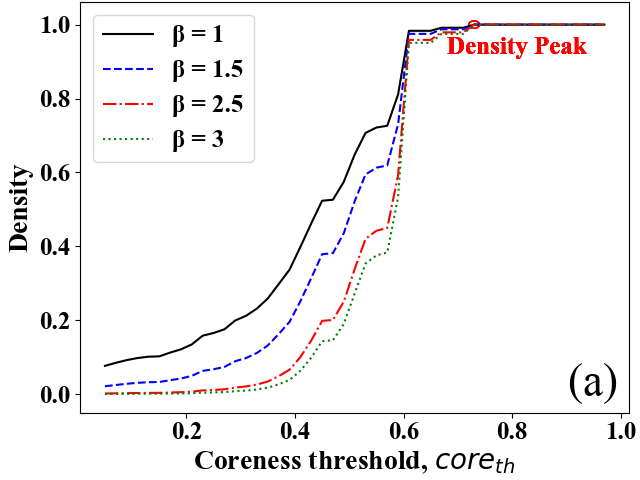} }}
    \subfloat[]{{\includegraphics[width=5.6cm]{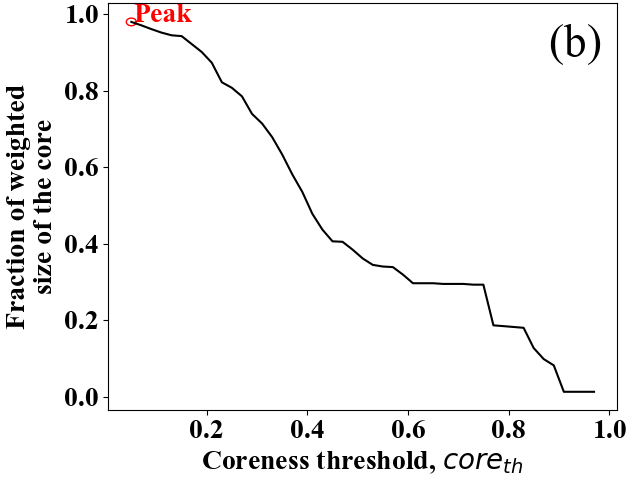} }}
    \subfloat[]{{\includegraphics[width=5.6cm]{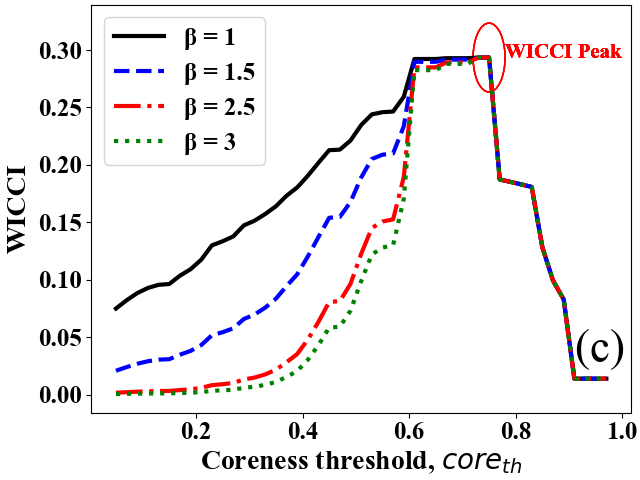} }} 
    \caption{Variation of (a) density, (b) fraction of weighted size of core, and (c) \wicci\ with varying  $core_{th}$. (a) Initially, the density dominates over fraction of weighted size of the core and hence \wicci~increases rapidly in (c). (b) In the later stages, the inverse happens -- fraction of weighted size of core dominates which results in \wicci~declining steeply in (c). The \wicci\ peak of $0.294$ is observed at $core_{th} = 0.73$ in (c). All nodes with a weighted coreness above  $core_{th}$ are part of the core. \textcolor{black}{Note that despite varying the density of core w.r.t \wicci~by changing the density coefficient ($\beta$), we observe a similar \wicci~peak in all cases.}
    }\label{fig:cii_vs_ct}%
\end{figure*}

\begin{algorithm}[!t]
\SetAlgoLined
 \nonl\textbf{Input:} \graphname\ $G(N,E)$ \\
 \nonl\textbf{Output:} $G_c$: Subgraph containing core nodes\\
 
 Initialize $wicci_{max}$ $\leftarrow$ 0\;

 \Comment{\textit{Running the weighted $k$-core decomposition on $G(N,E)$.}}
 
 $wc$ = List of weighted coreness scores for nodes in $G(N,E)$.
 
 \Comment{\textit{Sort N by $wc$ and push into stack $S$}}
 $\mathcal{S}$ = Stack of nodes in $G$ in descending order of weighted coreness, $wc$.

\Comment{\textit{Running set of core nodes}}
 $core_{n}$ $\leftarrow$ [\ ]

  \Comment{\textit{Set coreness threshold as the max weighted coreness}}
 $core_{th} \leftarrow max(wc)$

 \While{$core_{th}$ $>$ 0}{
    \Comment{\textit{Get node with maximum coreness}}
    $n = \mathcal{S}.pop()$\

    \While{$wc(n) >= core_{th}$}{
        \Comment{\textit{Add $n$ to  $core_{n}$}}
        $core_{n}.add(n)$\
        
        $n = \mathcal{S}.pop()$\
    }
    \Comment{\textit{As $wc(n) < core_{th}$, we push $n$ back to $\mathcal{S}$}}
    
  $\mathcal{S}.push(n)$\
   
    \Comment{\textit{Make induced subgraph of core using current $core_{v}$}}
    
    $G_{cr}$ $\leftarrow$   
    InducedSubgraph(G, $core_{n}$)\
    
    \Comment{\textit{Compute \wicci~for the current core $G_{cr}$}}
    
    wicci $\leftarrow$ WICCI($G_{cr}$,G)\
    
    \Comment{\textit{Finding the $G_{cr}$ with maximum \wicci}}
    \If{wicci $>$ $wicci_{max}$}{
        $wicci_{max}$ $\leftarrow$ wicci\\\
        $G_c$ = $G_{cr}$\
    }
\Comment{\textit{Iteratively decrease the coreness threshold}}
 $core_{th} \leftarrow core_{th} - 1$ 
 }

\caption{\algonamegraph~algorithm}
\end{algorithm}

\subsection{\algonamegraph: A Graph-based Method for Core Detection}
By considering the above properties of collusive entities, we design \algonamegraph\  (\textbf{K}-core decomposition for c\textbf{OR}e collu\textbf{S}ive us\textbf{E}rs detection), a modified version of (weighted) $k$-core decomposition  that is designed for detecting core users in blackmarket services based only on the topological structure of \graphname. 
It takes \graphname~as input and detects core blackmarket users (core subgraph $G_C$). \algonamegraph~is implemented by decreasing the coreness threshold and consequently making larger subgraphs of the core. The subgraph with the largest \wicci\ is our final core. 

Algorithm 1 presents the pseudo-code of \algonamegraph. Firstly, we apply weighted $k$-core decomposition which gives the weighted coreness score $wc(n)$ for each vertex $n \in N$. The vertices are then sorted in decreasing order of $wc$ and pushed into a stack $\mathcal{S}$. The top of the stack is the node with the maximum weighted coreness. Next, we create a running set ($core_n$) of core nodes initially with no node. The running coreness threshold $core_{th}$ is set to the maximum value of weighted coreness $wc_{max}$. Next, $core_{th}$ is iteratively decreased, and the set of core nodes is updated by adding all nodes $n$ which have $wc(n)$ greater than $core_{th}$. Next, $G_{cr}$, an induced subgraph is created  only by the core nodes. Further, \wicci~of $G_{cr}$ is calculated. The induced subgraph with the maximum \wicci~($wicci_{max}$) is the core of the graph, and the corresponding $core_{th}$ is the coreness threshold.

On applying \algonamegraph~on \graphname, we obtain an ideal coreness threshold of $0.73$ on a max-normalized scale, with a peak \wicci~ value of $0.294$ (c.f. Fig. \ref{fig:cii_vs_ct}) for different values of the density coefficient $\beta$. We explore the variation of \wicci~with $core_{th}$:
\begin{enumerate}

    \item Initially, as $core_{th}$ increases ($0.1-0.5$), users of low $wc$ (which contribute less to the overall collusive activity of the network) are removed from the core subgraph,  leading to rapid increase in density of the core subgraph and a relatively smaller decrease in the fraction of weighted size of the core. Initially, density dominates the fraction of weighted size of the core and hence \wicci~increases (c.f. Fig. \ref{fig:cii_vs_ct}(a)).
    
    \item Towards the higher values of $core_{th}$ ($>0.8$), density obtains its maximum value of $1$. However, the fraction of weighted size of the core decreases rapidly due to the continued exclusion of more nodes with relatively higher $wc$. As $core_{th}$ increases further, the fraction of weighted size of the core dominates density towards the latter values of $core_{th}$, and hence \wicci~decreases (c.f. Fig. \ref{fig:cii_vs_ct}(b)).
    
    \item In the mid-range values ($0.6- 0.7$) of $core_{th}$, the peak  of \wicci~is observed.  The corresponding core formed by the nodes (with $wc$ higher than $core_{th}$) leverages both the density and the fraction of weighted size of \graphname\ (c.f. Fig. \ref{fig:cii_vs_ct}(c)).
\end{enumerate}

\begin{figure*}[!t]
\captionsetup[subfigure]{labelformat=empty}
    \centering
    \subfloat[]{{\includegraphics[width=0.9\textwidth]{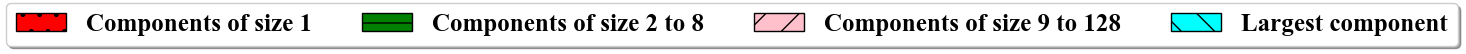} }} \\ 
    \subfloat[(a) Weighted degree]{{\includegraphics[width=3.8cm]{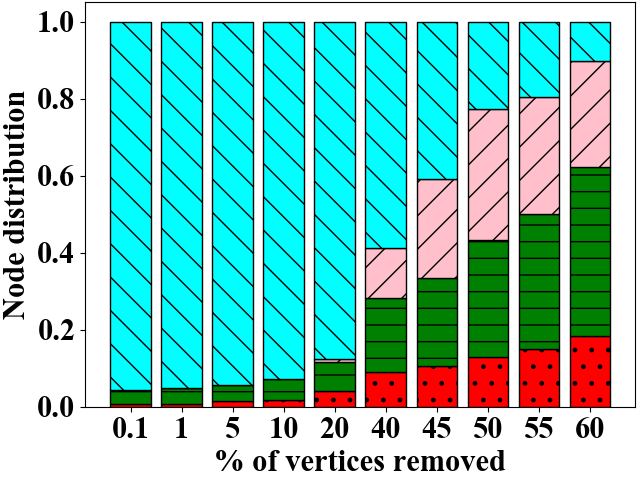} }}
    \subfloat[(b) Unweighted degree]{{\includegraphics[width=3.9cm]{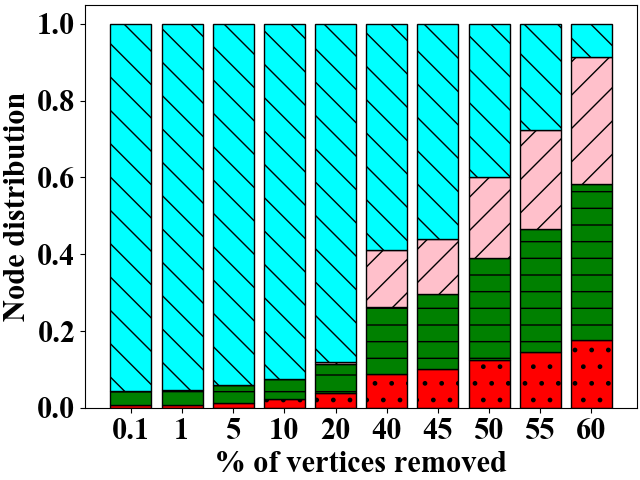} }}
    \subfloat[(c) Weighted coreness]{{\includegraphics[width=3.95cm]{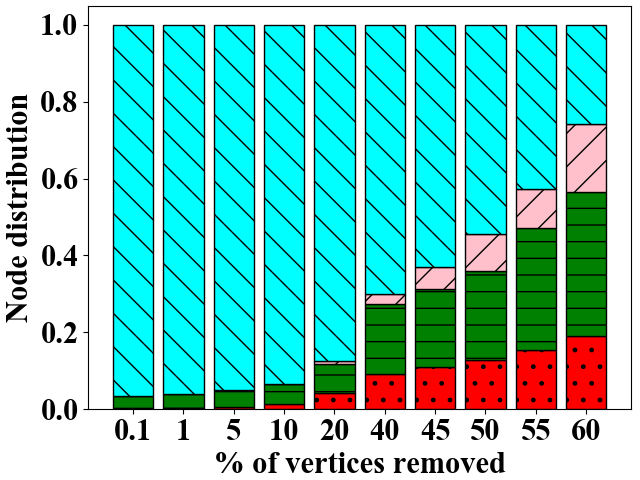} }}
    \subfloat[(d) Unweighted coreness]{{\includegraphics[width=3.95cm]{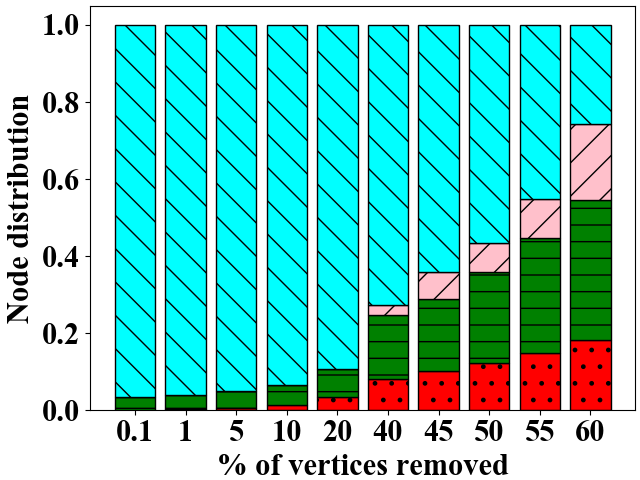} }}
    
    \caption{The distribution of nodes in components of sizes present in \graphname\ after removing nodes in the decreasing order of (a) weighted degree, (b) unweighted degree, (c) weighted coreness, and (d) unweighted coreness. The network visibly disintegrates into smaller components when at least (a) 50\% (b) 55\% (c) 60\%, and (d) 60\% are removed from the network. Despite a large removal of nodes, the remaining network has a high connectivity.  }
    \label{fig:breakage}%
    
\end{figure*}

The core obtained on applying \algonamegraph\ consists of $148$ nodes and (surprisingly) {\em is a complete graph}. Nearly $30\%$ of the entire collusive commenting activities of the network happens among $10\%$ of the core nodes. The periphery consists of $1,455$ nodes and has an edge density of $0.0355$. Nearly $60\%$ of the commenting activities take place among the peripheral nodes despite $90\%$ of the users belonging to it. The rest $10\%$ activities are captured between the core and the peripheral nodes (cross-edges between core and periphery). We now investigate the connectivity of the core in our proposed CCN network.
  






\section{Impact of Core on \graphname}\label{sec:impact_core_ccn}
To closely explore the connectivity of the core in the network, we analyse the effect after removing the core from \graphname. \citet{mislove2007measurement} reported that in a conventional social network, the removal of core breaks the graph into small disconnected components. However, in our case we notice that the graph does not break into smaller components even after removing a large fraction of core nodes (c.f. Fig.  \ref{fig:breakage}). The possible reasons for such a behavior are as follows:
\begin{enumerate}

\item \textbf{Estimated core may be incorrect:} One may argue that our metric \wicci~to estimate the core may be flawed. It may be possible that the core is larger than what we estimate. To verify this, we start by removing the vertices from \graphname~in the decreasing order of the (i) weighted degree (c.f. Fig. \ref{fig:breakage}(a)), and (ii) weighted coreness $wc$ (c.f. Fig. \ref{fig:breakage}(c)). We observe that the point where the size of the largest connected component decreases and the number of small disconnected components increases drastically, should be the appropriate value of $core_{th}$. However, we notice that such a point arises only after removing 50\% and 60\% of nodes based on weighted degree and weighted coreness of vertices from \graphname, respectively. This would suggest that at least 50\% of the vertices belong to the core. However, the density of the core reduces significantly (c.f. Fig. \ref{fig:density}). This violates one of the fundamental properties of a core that it should be incredibly dense. \citet{mislove2007measurement} observed near-complete degradation of the largest connected component after only removing $10\%$ of the nodes based on degree. Therefore, the observed pattern is not the artifact of our proposed metric \wicci, but a result of the high connectivity even among users of low coreness.
 
 \begin{figure}[!t]
\centering
\includegraphics[width=0.99\columnwidth]{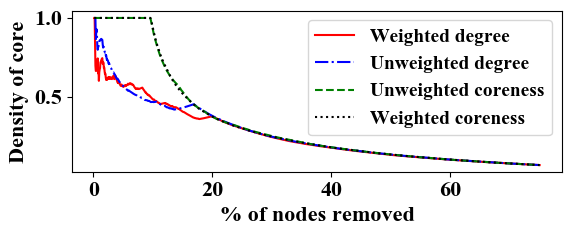}

\caption{Change in the density of core with the number of nodes removed.}\label{fig:density}

\end{figure}


    \item \textbf{\underline{Weighted} $k$-core decomposition may be incorrect:}
    One may argue that we should consider the traditional {\em unweighted} $k$-core decomposition \cite{mislove2007measurement}, instead of considering the weighted edges. 
    We perform similar experiments by removing vertices in the order of the (i) unweighted degree (c.f. Fig. \ref{fig:breakage}(b)) (as suggested in \citet{mislove2007measurement}) and (ii) unweighted coreness (c.f. Fig. \ref{fig:breakage}(d)). We observe similar results in both the cases where the network breaks into many small disconnected components upon removing at least 55\% of the nodes. This would again make the core incredibly sparse (c.f. Fig. \ref{fig:density}). Therefore, applying weighted $k$-core decomposition is not a reason for the late disintegration of the graph into smaller components.
\end{enumerate}
    \textbf{Possible explanation: connected periphery.} 
    We examine $G_P$, the induced subgraph of the peripheral nodes  independently with specific focus on its largest connected component $G_P^L$.
    
    \begin{itemize}
        \item $G_P^L$ and $G_P$ have $1,376$ and $1,455$ nodes, respectively.
        \item $G_P^L$ and $G_P$ have edge density of $0.03674$ and $0.0355$, respectively.
        \item $G_P^L$ has an average path length of $2.6355$.
        \item Lastly, as stated earlier, when we progressively remove the core from \graphname, the periphery largely remains intact.
    \end{itemize}
    
    This indicates that there is a significant connectivity among nodes in the periphery. This does not fall within the conventional structure of the periphery which  is generally described as small disconnected components. Instead, we visualize the periphery in $G_P^L$ as smaller and relatively dense communities (c.f. Fig. \ref{fig:demo}). \textcolor{black}{One possible reason for a  connected periphery may be that the graph has organically grown to a stage where despite the detection of the core users, the blackmarket service is in a self-sustainable stage and is no longer driven by the core users alone.
    A solution would be to detect the core users at an early stage to halt the growth of the market. To identify the network at its infancy, one would have to create multiple snapshots of the blackmarket services over a period of time, which is a computationally expensive task. We now examine the relation between the core and peripheral communities present in the proposed network.}

\section{Interplay Between Core and Peripheral Communities}
Here, we study the interactions between the core and periphery, and highlight critical observations. We start by dividing the videos $V$ into  three categories:
\begin{enumerate}
    \item \textbf{Core-core videos} are the set of videos commented exclusively by core users.
    \item \textbf{Core-periphery videos} are the set of videos commented by both core and peripheral users.
    \item \textbf{Periphery-periphery videos} are the set of videos commented  exclusively by peripheral users.
\end{enumerate}
Here, (1) and (2) are responsible for the formation of edges within core;
(2) and (3) are responsible for the formation of edges within periphery;
(2) alone is responsible for the formation of edges between core and periphery.

Next, we  define the community structure in \graphname. A ``good'' community in \graphname~is the one in which the users of the community have co-commented heavily on a set of videos. Due to the high connectivity observed in the periphery (mentioned in the earlier section),
we speculate that the periphery consists of several small communities. To check this, we run the weighted version of the Louvain community detection method \cite{blondel2008fast} for detecting peripheral communities $C^L_P$ from $G^L_P$ (the largest connected component in the induced subgraph in the periphery). The modularity of the community structure detected by Louvain is $0.397$, and the number of large communities (with size $>40$) is $9$. It indicates that there exist large communities of collusive users that comment on the same set of videos. Next, we define the interaction within the  peripheral community based on the amount of  collusive commenting activities occurring inside the community. We categorize these interactions using (a) weighted size, and (b) average weighted degree of nodes in the peripheral community. We also quantify the interactions between  core and each of the peripheral communities based on the amount of  commenting activities on the core-periphery videos. 

[\textbf{Internal Interaction of Peripheral Community}] 
We define the internal interaction of a peripheral community as a measure of the collusive commenting activities within the community.

\noindent We further categorize the internal interaction using the following metrics:

\begin{enumerate}
    \item \textit{\textbf{Average weighted degree of nodes in the community}}: It captures the average collusive commenting activities taking place within the community.
    
    \item \textit{\textbf{Weighted size of the community}}: It is measured by the sum of weights of all the internal edges of a community, capturing the total intra-community  collusive commenting activities.
    
\end{enumerate}

[\textbf{Independent Interaction of Core and Peripheral Community}]
We define the independent interactions of core and a peripheral community as a measure of the collusive commenting activities taking place between the core and the peripheral community. This indicates the participation of the peripheral users in commenting on core-periphery videos.

 To capture  independent interactions between core and peripheral community $C$, we utilize the {\bf weighted cut-set}  $WCS_{Core,C}$ as the sum of the weights of edges connecting the core and $C$. Since the size of the peripheral communities varies, we normalize $WCS_{Core,C}$ by only $|C|$.
 \begin{equation*}\label{eq:cut}
WCS_{core,C}=\frac{\textit{\small Sum of weights of edges connecting core and $C$}}{|C|}
\end{equation*}

\begin{figure}[!t]
    \captionsetup[subfigure]{labelformat=empty}
    \centering
    \subfloat[]{{\includegraphics[width=0.49\columnwidth]{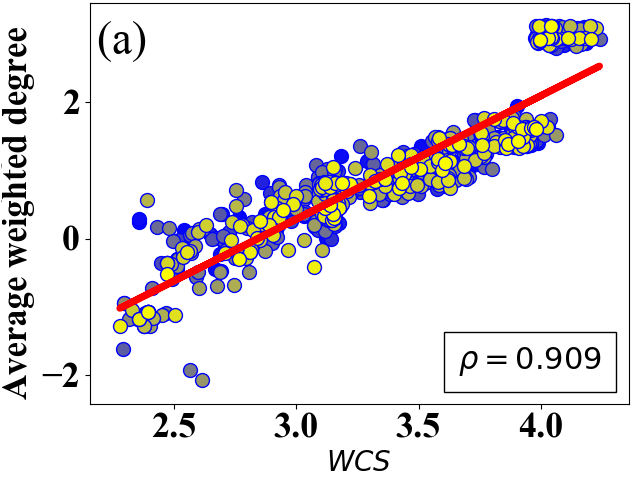}}}
    \subfloat[]{{\includegraphics[width=0.49\columnwidth]{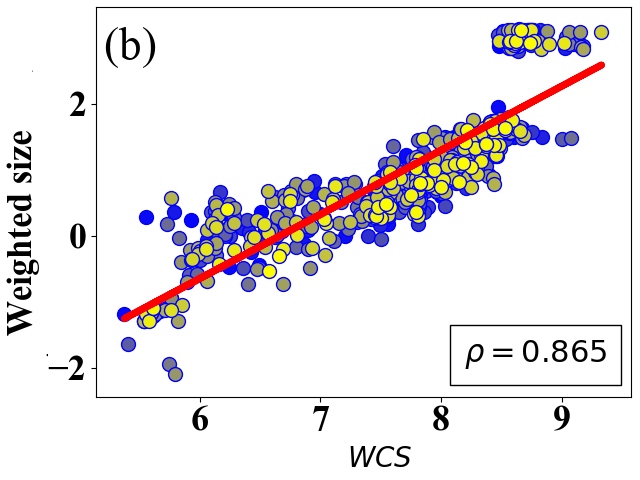}}}
    
    \caption{A strong positive correlation between weighted cut-set $WCS$ and -- (a) average weighted degree, and (b) weighted size of the peripheral communities. Different colors indicate communities obtained in different executions of Louvain method. The Pearson's $\rho$ is also reported. 
    }    \label{fig:interaction}
\end{figure}
 
\noindent The following observations are drawn from the above (c.f. Fig \ref{fig:interaction}):
\begin{enumerate}
    \item There exists a positive correlation between the average weighted degree of a peripheral community and $WCS_{core,C}$ (c.f. Fig \ref{fig:interaction}(a)).
    
    \item There exists a positive correlation between the weighted size of a peripheral community and $WCS_{core,C}$ (c.f. Fig \ref{fig:interaction}(b)).
    
\end{enumerate}

From these observations, we conclude that there is a definite positive correlation between the internal interaction within the peripheral communities and that between the core and peripheral communities. Peripheral communities which actively participate in activities associated with the core (such as commenting on core-periphery videos), tend to contribute more to the collusive market. We now discuss in our detail our proposed deep fusion framework \algonameml~for the identification of core blackmarket users.



\section{\algonameml: A Deep Fusion Framework} \label{sec:nurse}
Although the network topology based weighted $k$-core decomposition presented in \algonamegraph\ is highly accurate to detect core blackmarket users, it may not be feasible to adopt in designing a real-world system because of the following reasons: (i) data arrives in streaming fashion, and the generation of \graphname\ is not possible as the entire snapshot of the blackmarkets at a certain point is impossible to collect; (ii) \graphname\ is often incomplete and highly sparse, and (iii) $k$-core decomposition is comparatively slow. {\bf However, we consider \algonamegraph\ as an oracle and the core and compromised users it has detected as the ground-truth to train and evaluate the following model.} To address the above issues and towards designing a real-world system, we propose \algonameml\ (\textbf{N}e\textbf{U}ral framework for detecting co\textbf{R}e collu\textbf{S}ive us\textbf{E}rs), a neural fusion model to detect core blackmarket users in blackmarket services {\em based only on the user timeline and video sharing information} (without considering the underlying \graphname).

\subsection{\algonameml: Model Components}
\begin{figure}[!htbp]
\centering
\includegraphics[width=0.8\columnwidth]{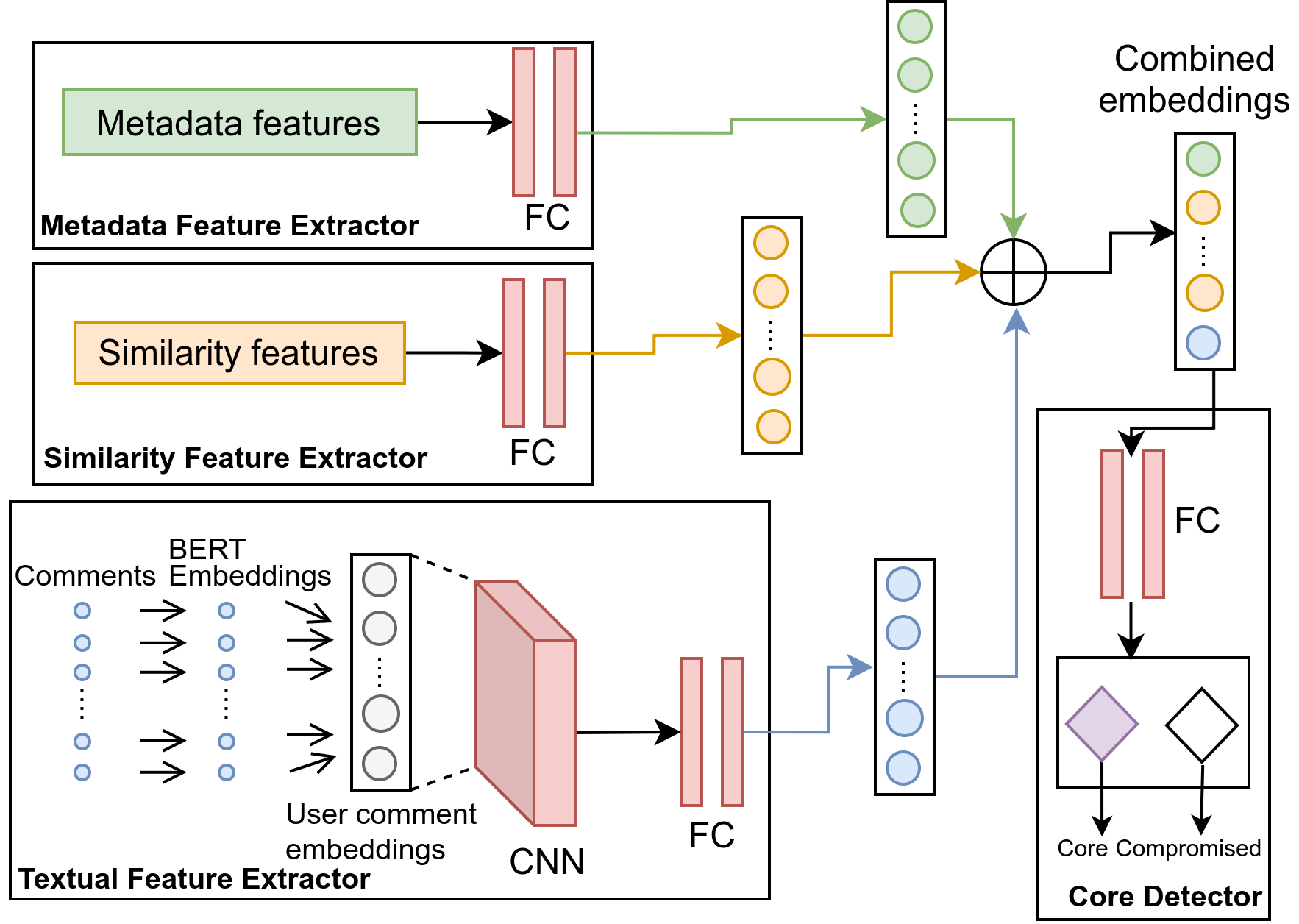}
\caption{A schematic diagram of \algonameml. The green colored network is the {\em metadata feature extractor} \textbf{(MFE)}, the orange colored network is the {\em similarity feature extractor} \textbf{(SFE)}, and the blue colored network is the {\em textual feature extractor} \textbf{(TFE)}. We concatenate the output of the feature extractors to form the feature representation of a YouTube user. The final representation is passed through to a \textit{core detector} module to detect whether the given user is a core user or a compromised user.}
\label{fig:architecture}

\end{figure}

 \algonameml\ comprises three components: {\em metadata feature extractor} \textbf{(MFE)}, {\em similarity feature extractor} \textbf{(SFE)}, and {\em textual feature extractor} \textbf{(TFE)}; the output of which are further concatenated to form the feature representation of a YouTube user. The combined representation is passed through to a \textit{core detector} module which determines whether the user is a core or a compromised user. The architectural diagram of \algonameml~ is shown in Fig. \ref{fig:architecture}. Individual components of \algonameml\ are elaborated below.


\subsubsection{Metadata Feature Extractor \textbf{(MFE)}.} 
We extract $26$ metadata features based on the profile information, and videos uploaded by the users. These features are largely divided into four categories:\\
\hspace*{2mm}(a) \textbf{Self-comments ($MFE_{1-5}$)}: These features are derived from the comments made by the users on their own videos. We observe that, on an average, compromised users tend to write more self-comments ($\times 1.778$) than the core users, indicating that {\em core users are less involved in self-promotion}. We take the maximum, minimum, total, average and variance of the comments across self-posted videos as five different features.\\
\hspace*{1.5mm}(b) \textbf{Number of videos uploaded ($MFE_{6}$)}: It refers to the  total number of videos uploaded by the user. On average, core users upload fewer videos, which is $\times 0.633$ less than that of compromised users. {\em A core user's efforts to benefit from the blackmarkets are lesser} (as they are created by the blackmarket services themselves) than the compromised  users.\\
\hspace*{1.5mm}(c) \textbf{Duration of uploaded videos ($MFE_{7-11}$)}: These features measure the duration of the videos uploaded by users. On average, a core user uploads significantly shorter videos, which is $\times 0.628$ less than that of compromised users. The possible reason could be that {\em core users  are less interested in their own content}; rather their primary objective is to artificially inflate the popularity of other customers' videos. We take the maximum, minimum, total, average and variance of video duration per user as five different features.\\
\hspace*{1.5mm}(d) \textbf{Other features:} Apart from the above features, we also consider the following features  related to the rating of the videos posted by a user (in each case, we take the maximum, minimum, total, average and variance as five different features) -- the number of likes {$(MFE_{12-16}$)}, the number of dislikes {$(MFE_{17-21}$)} and the number of views received {$(MFE_{22-26}$)}.

\subsubsection{Similarity Feature Extractor \textbf{(SFE)}.} 
Collusive users have been shown to post similar/duplicate comments regardless of the topic of the content \cite{dutta2020detecting}. We extract two sets of features  based on the linguistic similarity of comments posted on the video and video metadata:\\
\hspace*{2mm}(a) {\bf Comment-based features}:
We capture similarity features based on the linguistic similarity of comments posted by users. For a user, let the set of her comments on her own videos and the set of comments on other videos be $SC$ and $OC$, respectively. We first generate embedding of individual comments using pre-trained BERT \cite{devlin2018bert}. We then measure the maximum, minimum, total, average and variance of similarities (cosine similarity) between comments in $SC$. Similarly, we obtain five similar features, each from the comments within $OC$ and by comparing comments in $SC$ and $OC$. This results in $15$ features ($SFE_{1-15}$).\\
\hspace*{2mm}(b) {\bf Video metadata based features}: In YouTube, a user can upload her own videos ($SV$) or act on videos posted by other users ($OV$). For each video, we combine the text of the video title, video description and video genre. We then generate the embedding of the combined text using BERT. Next, we extract the maximum, minimum, total, average and variance of similarities (cosine similarity) between video embeddings, each from the videos within $SV$ and and videos across $SV$ and $OV$. This results in 10 features denoted by $SFE_{16-25}$. We did not extract features from within $OC$ because we observed that doing so heavily biased the model.

\subsubsection{Textual Feature Extractor \textbf{(TFE)}.} We capture textual features from the content of the comments posted by a user. We generate embeddings for every comment using pre-trained BERT \cite{devlin2018bert}. To get a representative  embedding for a user, we average out the embeddings of all the comments posted by the user. As collusive users tend to post repetitive text in their comments \cite{dutta2019blackmarket}, we feed the resultant embedding into a CNN to capture this inter-dependency. In literature, CNNs have shown to perform well in  capturing repetitive patterns in texts \cite{lettry2017repeated,zhou2018sentiment}.

\subsubsection{Core Detector.} The core detector module consists of a fully-connected layer (FC) with softmax to predict where a YouTube user is core or compromised, denoted by $G_c(.,\theta_c)$, where $\theta_c$ represents the model parameters. For the prediction task, $G_c$ generates the probability of a user $u$ being the core user based on the combined representation $\vec{u}$.
\begin{equation}
    P_{\theta}(u) = G_c(\vec{u};\theta_c)
\end{equation}

We use the cross-entropy loss ($L_d$) for our model:
\begin{equation}
    L_d(\theta) = y \log\big(P_{\theta}(u)\big) + (1-y)\log\big(1-P_{\theta}(u)\big)
\end{equation}


\subsection{\algonameml: Model Specifications} \algonameml\ executes three parallel operations - {\bf (1) TFE:} The $1\times784$ textual vector is fed to a CNN (number of channels =  $32$, filter size  = $2$, no padding). Next, the resultant vector is passed to a max-pooling layer and then to a FC layer of size $64$. The final output from this operation is a $1\times64$ vector.
{\bf (2) SFE:} The $1\times25$ similarity vector is fed to a FC Layer of size $32$. A dropout of $0.3$ is applied on the FC layer. The final output from this operation  is a  $1\times32$ vector.
{\bf (3) MFE:} The $1\times26$ metadata vector is passed to a FC layer of size $16$. A dropout of $0.25$ is applied on the FC layer. The final output from this operation is a $1\times16$ vector.

The combined representation is a $1\times112$ vector. This is then passed to another FC layer of size $16$, followed by a softmax layer of size $2$ to obtain the final prediction. We utilize the ReLu activation function for all other layers.



\section{Experiments}
\subsection{Dataset and Ground-truth} 
Although we collected collusive users from the blackmarkets, it is unknown who among them are core blackmarket users. Thus, the ground-truth information about the core and compromised users are impossible to obtain unless blackmarkets themselves provide the data! We, therefore, consider the core and compromised users obtained from \algonamegraph\ as the ground-truth since it uses the topological structure of the underlying collusive network to detect the core users. We hypothesize that \algonamegraph\ is highly accurate in detecting core users. We also perform several case studies to validate our hypothesis. We intend to show how much \algonameml\ (a non-topology based method) is close to \algonamegraph\ (a pure topology-based method). We also present a case study to show whether the detected core users are really meaningful or not. 

Since the number of compromised users ($1,455$) is $~$$10$ times higher than the number of core users ($148$), we generate two datasets for our analysis: {\bf (i) Dataset (1:1)} is a balanced dataset where equal number  of compromised users as that of core users are (randomly) sampled; 
{\bf (ii) Complete dataset} is an imbalanced dataset where all collusive users are kept. 
We performe $10$-fold stratified cross-validation and report the average performance.
\subsection{Baseline Methods}
Since ours is the first work to detect core blackmarket users, there is no existing baseline. We therefore design our own baselines by considering individual components of \algonameml\ in isolation and their combinations:
\begin{enumerate}
    \item \textbf{MFE:} This model uses only the metadata feature extractor.
    
    \item \textbf{SFE:} This model  uses only the similarity feature extractor.
    
    \item \textbf{TFE:} This model uses only the textual feature extractor. Each comment is represented as a $786$ dimensional vector using BERT.
\end{enumerate}

We further combine these three components and design three more baselines: (4) \textbf{MFE+SFE}, (5) {\bf MFE+TFE}, and (6) {\bf SFE+TFE}.

These baselines also in turn serve the purpose of {\bf feature ablation} to explain which features are important for \algonameml.

{\em Are core users the influential nodes in the network}? To answer this, we consider three other approaches as baselines which aim to detect influential users:
\begin{enumerate}
\setcounter{enumi}{6}
    
  \item {\bf INF:}  \citet{huang2020identifying} proposed a node influence indicator, called INF, based on the local neighboring information to detect influential nodes in a network. 
  \item {\bf Weighted Betweenness  Centrality (WBC):} 
  Betweenness centrality (BC) \cite{brandes2001faster} is a measure of node centrality based on the shortest paths. We utilize the approach in \cite{shin2016corescope} to run the weighted version of BC on \graphname\ and detect core users.
  \item {\bf Coordination Game Model (CGM):} \citet{zhang2017top} proposed a coordination game model to find top-$K$ nodes to maximize influence under certain spreading model.
\end{enumerate}

\begin{figure}[!t]
    \centering
    
    \captionsetup[subfigure]{labelformat=empty}
    \subfloat[]{{\includegraphics[width=0.49\columnwidth]{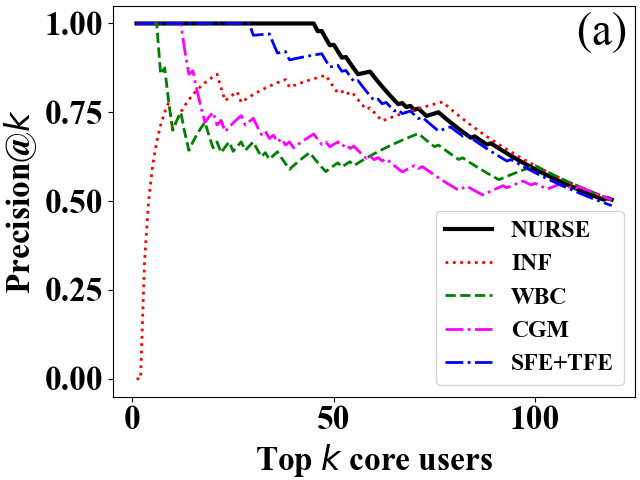}}}
    \subfloat[]{{\includegraphics[width=0.49\columnwidth]{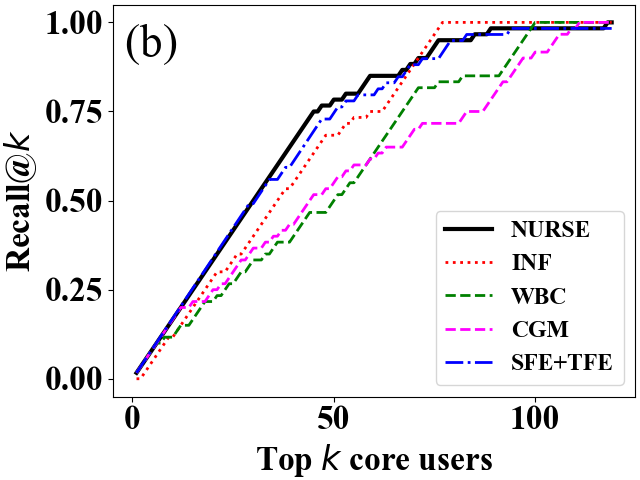}}}
    
    \caption{Change in the performance of competing methods with the increase of $k$ (the number of results returned) for detecting core users from our dataset (1:1). For better visualization, among the variations of \algonameml, we report the results of only the best variation (SFE+TFE).} 
    \label{fig:performance}
\end{figure}

\begin{table}[!htbp]
\centering
\caption{Performance (F1-Score and AUC for detecting core users) of the competing methods at $k=148$ (break-even point). The results also explain \underline{feature ablation} of \algonameml.}\label{table:performance}

\begin{tabular}{|l||c|c||c|c|}
\hline
{{\bf Method}} & \multicolumn{2}{c||}{{\bf Dataset ($1:1$)}} & \multicolumn{2}{c|}{{\bf Complete Dataset}} \\\cline{2-5}
& {\bf F1 (Core)} & {\bf AUC} & {\bf F1 (Core)} & {\bf AUC} \\\hline
MFE &0.638  & 0.559 & 0.268  & 0.294 \\
SFE &0.816  & 0.857  & 0.516  & 0.472 \\
TFE &0.665  & 0.773 & 0.530 & 0.365 \\\hline
MFE+SFE & 0.824 & 0.882 
 & 0.682 &0.718\\
MFE + TFE &0.696 &0.767 &0.415 &0.631\\
SFE+TFE & 0.819 & 0.865 &0.721 & 0.792\\\hline
INF &0.750 &0.139 &0.533 & 0.113\\
WBC &0.617 &0.304 &0.407 &0.270\\
CGM &0.622 &0.392 &0.302 &0.414 \\\hline
\algonameml & \textbf{0.879}
 & \textbf{0.928} & \textbf{0.833} & \textbf{0.845}\\\hline
\end{tabular}
\end{table}

\subsection{Performance Comparison}
Since all the competing methods return a score (or a probability), indicating the likelihood of a user being core, we first rank all the users based on the decreasing order of the score, and then measure the accuracy in terms of precision, recall, F1-Score and Area under the ROC curve (AUC) w.r.t. the `core' class. Fig. \ref{fig:performance} shows that \algonameml\ dominates other baselines for almost all values of $k$ (the top $k$ users returned from the ranked list). Table \ref{table:performance} summarizes the performance (F1-Score and AUC) of the models at $k=148$ (as there are $148$ core users; it is also known as break even point) -- \algonameml\ turns out to be the best method, followed by MFE+SFE (for balanced dataset) and SFE+TFE (for imbalanced dataset). Similarity feature extractor (SFE) seems to be the most important component of \algonameml, followed by TFE and MFE. Among influential node detection methods, both INF and CGM seem to be quite competitive. Next, we examine the core users identified by our proposed method \algonamegraph~and \algonameml.


\section{Case Studies}\label{sec:case:study}
 We further delve deeper into the characteristics of some of the core users detected by both \algonamegraph\ and \algonameml\ by conducting some case studies. These provide us strong evidences to validate our strategy of collecting the ground-truth from \algonamegraph.
\begin{enumerate}
    
    \item \textbf{Core users are heavy contributors}: A core user, on average, comments significantly \textbf{($\times2.665$)} more than a compromised user, indicating that core users are the top contributors to the freemium collusive market. 
    
    \item \textbf{Despite being heavy contributors, core users are not the  largest beneficiaries of the collusive market}: 
    We measure the average number of comments received by the videos uploaded  by  collusive users, and rank them in decreasing order of this quantity.  We find only one core user from the top $30$ users. Upon further investigation, we notice that only $8$ out of the top $250$ users are core users. This suggests that core users, despite being heavy contributors, are not the largest beneficiaries of the collusive market. 
    
    \item \textbf{Core users aggressively  participate in the collusive market}: We observe that the average number of comments made per collusive video by core users is twice \textbf{$(\times1.997)$} higher than that of compromised users. This indicates an aggressive behavior to promote the videos they comment on. 
    \item \textbf{Channels controlled by core users are not popular}:
    We observe that the channels controlled by core users are not the popular YouTube channels. More than $85\%$ of the channels have a subscriber count of less than $1,000$. This clearly indicates that the primary objective of the core users is not to promote their own videos/channels.
    
    \item \textbf{Channels controlled by core users have less uploaded videos}:
    We observe that the channels controlled by core users usually do not contain much YouTube videos. More than $90\%$ of the channels have a video count of less than $100$. This further corroborates the theory behind the working principle of core blackmarket users. 
\end{enumerate}

Despite the above suspicious characteristics exhibited by core channels, we observe that till date, $~$$93\%$ of the core channels \textbf{continue to be active on YouTube}. On average, these core channels have been active on YouTube for over $4$ years ($1497$ days).
It indicates how core channels are able to evade the current in-house fake detection algorithms deployed by YouTube.
\color{black}

\section{Conclusion}
This paper addressed the problem of detecting core users in YouTube blackmarkets. We curated a new dataset of collusive YouTube users. We then proposed \algonamegraph, a novel graph-based method to segregate core users from compromised accounts. Empirical studies revealed interesting dynamics of core and compromised users. As \algonamegraph\ is practically infeasible to design due to its dependency on the underlying collusive network,  we further proposed \algonameml, a deep fusion model that leverages only the user timeline and video submission information to detect core users. Extensive experiments on our dataset showed that \algonameml~is highly similar to \algonamegraph\ in detecting core users. Summarizing, our study contributed in four aspects -- problem definition, dataset, methods and empirical observation. As a future work, it would be interesting to see how \texttt{NURSE} can merge with existing collusive entity detection approaches  to effectively identify core, collusive and non-collusive users. We also made the code and dataset available for the purpose of reproducibility.

\bibliographystyle{aaai-named}
\bibliography{references}

\end{document}